# Bacterial cooperation leads to heteroresistance

Shilian Xu[1,*,+], Jiaru Yang[2,+], Chong Yin[3,+]

[1]Department of Microbiology, Biomedicine Discovery Institute, Monash University, Melbourne, VIC, Australia, 3800

[2]Department of Biochemistry and Molecular Biology, Biomedicine Discovery Institute, Monash University, Clayton, VIC, Australia, 3800

[3]Bone Metabolism Lab, Key Laboratory for Space & Biotechnology, Institute of Special Environment Biophysics, School of Life, Northwestern Polytechnical University, Xi'an, Shanxi, China

[*]The corresponding author: nashduan@gmail.com, Shilian.Xu@monash.edu

[+]All authors contribute equally to this work

**Abstract:** By challenging E. coli with sublethal norfloxacin for 10 days, Henry Lee and James Collins suggests the bacterial altruism leads to the population-wide resistance. By detailedly analyzing experiment data, we suggest that bacterial cooperation leads to population-wide resistance under norfloxacin pressure and simultaneously propose the bacteria shield is the possible feedback mechanism of less resistant bacteria. The bacteria shield is that the less resistant bacteria sacrifice the large number of themselves to consume norfloxacin and then to relieve the norfloxacin burden from highly resistant bacteria. Moreover, by employing Game Theory, the interaction between highly resistant bacteria and less resistant bacteria under norfloxacin can be considered as Iterated Prisoner's Dilemma and the best strategy is to cooperate with each other, which agrees with bacterial cooperation achieved by analyzing experiment data. Thus, due to highly resistant bacteria and less resistant bacteria extracted from the same bacteria population, bacterial cooperation leads to heteroresistance. Additionally, the genetic relatedness of highly resistant bacteria and less resistant bacteria are very close, but the bacteria cooperate with each other, which provides a contradiction against Kin Selection Theory.

## Introduction

The extensive and intensive employ of antibiotics for clinical, agricultural and pharmaceutical purpose is forming and transforming sublethal bactericide pressure toward bacteria, which triggers multidrug resistance via radical-induced mutagenesis [1]. Moreover, the susceptibility of bacteria to bactericides may not be uniform in the whole bacterial population. Some bacteria in a population may remain susceptible to the bactericide, whereas other bacteria display varying degrees of drug resistance, which is known as heteroresistance [1]. In fact, the bacteria with different resistance levels are genetically different [2].

Henry Lee and James Collins challenged E.coli (MG 1665) by sublethal concentrations of norfloxacin and gentamicin and suggested that highly resistant isolates (HRIs) release signaling molecular, indole, to protect less resistant isolates (LRIs). Moreover, the protection provided by highly resistant isolates (HRIs) for less resistant isolates (LRIs) can be considered as the altruistic behaviors, which is known the form of kin selection [2,3]. However, the data mentioned by Lee at al might not suggest the major conclusion that HRIs is altruistic to LRIs under sublethal norfloxacin and gentamicin pressure, but support alterative and valid conclusion that HRIs and LRIs cooperate with each other under norfloxacin pressure [2,3].

In this manuscript, we would like to illustrate bacterial cooperation leads to heteroresistance under sublethal norfloxacin pressure by analyzing experiment data and by employing Game Theory. Moreover, this case may be contradiction to Kin Selection Theory.

**The experiment data indicates that bacterial cooperation enhances population-wide resistance and bacteria shield is possible feedback mechanism of LRIs**

First, by detailedly analyzing Henry Lee and James Collins's experiment data, the two major reasons for conclusion that HRIs and LRIs cooperate with each other is as follows.

1. **The abnormal changes of the proportion of HRIs and LRIs in co-culture**

The initial proportion of HRIs and LRIs is 1:100, but after 15-18 hours culture, the finial HRI-and-LRI ratio is 23:77. According to Table.1[5], if HRIs are altruistic to LRIs, the HRI-and-LRI ratio could be less than 1:100, because HRIs cost their fitness to protect LRIs. Then, the possible final HRI-and-LRI ratio could be 1:100 or even 1:1000. **Thus, this contradicts with the statement that HRIs is altruistic to LRIs.**

2. **The abnormal changes of the population of HRIs and LRIs in the isolation and co-culture**.

All norfloxacin concentrations are 1500 ng/ml. All colonies are cultured overnight (15-18 hours) and all initial populations of bacteria are $10^8$ cfu/ml, for both isolation and co-culture. In the isolation, only HRIs or LRIs is cultured. In the co-culture, the initial ratio HRI-to-LRI ratio is 1:100 (1 in 100). Namely, the initial populations of HRLs and LRIs are $10^5$cfu/ml and $10^7$cfu/ml.

2.1. The analysis for LRIs in the isolation and co-culture

In the isolation, under the norfloxacin pressure, the population of LRIs changes from $10^8$cfu/ml to $10^2$cfu/ml, which lowers $10^6$ times. In co-culture, the population of LRIs changes from $10^7$cfu/ml to $10^4$cfl/ml, which lowers $10^3$ times. This indicates that **the existence of HRIs improves the survival of the LRIs in co-culture**, which coincides with Lee's conclusion.

2.2. The analysis for HRIs in the isolation and co-culture

In the isolation, under then norfloxacin concentration 1500 ng/ml, the population of HRIs changes from $10^8$cfu/ml to $10^4$cfu/ml, which lowers $10^4$ times. In co-culture, the population of HRIs changes from $10^5$cfu/ml to $10^4$cfu/ml, which lowers $10^1$ times. This indicates that **the existence of LRIs improves the survival of the HRIs under norfloxacin pressure, although it**

**maybe contradicts against common sense**. **Moreover, this conclusion challenges the Lee's result.**

Here, according to Table 1 [5], it is easy to arrive that **HRIs and LRIs cooperate with each other under norfloxacin pressure**. Meanwhile, by comparing with the LRIs populations in the isolate and co-culture, the LRIs sacrifice the large number of themselves to protect HRIs in co-culture. Namely, the feedback mechanism of LRIs could be considered as **"the bacteria shield"** for the HRIs under norfloxacin pressure. Moreover, because the HRIs and LRIs are extracted from the same bacteria population challenged by norfloxacin for 10 days and exhibiting high levels of resistant heterogeneity, the cooperation among subpopulations in bacteria can be a valid mechanism for heteroresistance.

Table 1. A classification of social behaviors [5]

|  |  | Effect on recipient | |
|---|---|---|---|
|  |  | Positive | Negative |
| Effect on actor | Positive | Mutual benefit (Cooperation) | Selfishness |
|  | Negative | Altruism | Spite |

**The Iterated Prisoner's Dilemma suggests mutual cooperation is the best strategy for Highly Resistant Isolates and Less Resistant Isolates under norfloxacin pressure.**

In the section, we would like first to introduce several essential concepts for following analysis. The definition of Prisoner's Dilemma, Iterated Prisoner's Dilemma, Cooperation, Altruism and Kin Selection Theory.

Prisoner's Dilemma represents two members of a criminal gang are arrested and imprisoned. Each prisoner is in solitary confinement with no means of communicating with the other. The prosecutors lack sufficient evidence to convict the pair on the principal charge. They hope to get both sentenced to a year in prison on a lesser charge. Simultaneously, the prosecutors offer each prisoner a bargain. Each prisoner is given the opportunity either to: betray (defect) the other by testifying that the other committed the crime, or to cooperate with the other by remaining silent. The offer is:

a) If A and B each betray the other, each of them serves 2 years in prison;
b) If A betrays B but B remains silent, A will be set free and B will serve 3 years in prison (and vice versa);
c) If A and B both remain silent, both of them will only serve 1 year in prison (on the lesser charge).

Thus, in this case, the best strategy for each player is mutual defection [4].

The conditions of Iterated Prisoner's Dilemma are exactly same with those of Prisoner's Dilemma, but the game round is many times (repeated/ iterated), rather than once. In this case, the best strategy for each player is tit for tat. Namely, mutual defection and cooperation are the best strategy. Thus, cooperation can be formed [4].

Cooperation is a form of working together in which one individual pays a cost and another gains a benefit [4]. Altruism is the behavior that reduces fitness of actors and their offspring but efficiently increases fitness of recipients and their offspring [4]. Kin-selection theory suggests that the more similar genetic relatedness between actors and recipients, the more possible altruistic behaviors occur from actors to recipients [4].

From the perspective to Game Theory, we consider highly resistant isolates (HRIs) and less resistant isolates (LRIs) are considered as two prisoners. Moreover, norfloxacin pressure is regarded as offer provided by prosecutors. Because the generation time of E.coli is approximately 30 minutes, the culture time 15-18 hours can be considered as iterated game rounds and long-time evolution. According to Iterated Prisoner's Dilemma [4], the best strategy for both highly resistant isolates (HRIs) and less resistant isolates (LRIs) is to cooperate with each other, which coincides with the bacterial cooperation achieved by analyzing experiment data.

Moreover, by employing whole-genome sequence, Henry Lee and James J Collins show that LRIs and HRIs are very genetically closed. According to Kin Selection Theory, HRIs is supposed to be altruistic to LRIs or vice versa, which agrees with Henry Lee and James J Collins' conclusion. However, due to bacterial cooperation achieved by analyzing experiment data and employing Game Theory, particularly Iterated Prisoner's Dilemma, Kin Selection Theory and its corresponding conclusion that bacterial altruism leads to population-wide resistance may not hold.

**Conclusion**

In this manuscript, we suggest that bacterial cooperation leads to population-wide resistance under norfloxacin pressure and simultaneously propose the bacteria shield is the possible feedback mechanism of less resistant bacteria. The bacteria shield is that the less resistant bacteria sacrifice the large number of themselves to consume norfloxacin and then to relieve the bactericide burden from highly resistant bacteria. Moreover, by employing Game Theory, the interaction between highly resistant bacteria and less resistant bacteria under norfloxacin can be considered as Iterated Prisoner's Dilemma and the best strategy is to cooperate with each other, which agrees with bacterial cooperation achieved by analyzing experiment data. Thus, due to highly resistant bacteria and less resistant bacteria extracted from the same bacteria population, bacterial cooperation leads to heteroresistance. Additionally, the genetic relatedness of highly resistant bacteria and less resistant bacteria are very close, but the bacteria cooperate with each other, which provides a contradiction against Kin Selection Theory.